\documentclass[conference]{IEEEtran}
\IEEEoverridecommandlockouts
\usepackage{booktabs}
\usepackage{colortbl}
\usepackage{color}
\usepackage{xspace}
\usepackage{url}
\usepackage{balance}
\usepackage{tabulary}
\usepackage{array}
\usepackage{graphicx}
\usepackage[ruled,lined,linesnumbered,vlined,algo2e]{algorithm2e}
\usepackage{subfig}
\usepackage{listings}
\usepackage{lipsum}
\usepackage{hyperref}
\usepackage{graphicx}
\usepackage{amsmath,amssymb}
\usepackage[T1]{fontenc}
\usepackage{amsmath, amsthm, amssymb}
\usepackage[ansinew]{inputenc}
\usepackage{xcolor}
\usepackage{extarrows}
\usepackage{booktabs}
\usepackage{multirow}
\usepackage{colortbl}
\usepackage{tabularx}
\usepackage{enumitem}
\usepackage{ragged2e}
\usepackage{balance}
\usepackage[font=footnotesize,labelfont=bf]{caption}
 \def\BibTeX{{\rm B\kern-.05em{\sc i\kern-.025em b}\kern-.08em
    T\kern-.1667em\lower.7ex\hbox{E}\kern-.125emX}}

\usepackage{algorithm}
\usepackage{algorithmic}

\definecolor{myred}{rgb}{1,0,0}
\definecolor{myblue}{rgb}{0,0,1}
\definecolor{mymauve}{rgb}{0.58,0,0.82}

\newcommand{\mytool}{\textsc{HPDroid}\xspace}
\usepackage{pifont}

\begin{document}

\title{An Empirical Evaluation of GDPR Compliance Violations in Android mHealth Apps}

\author{Ming Fan\IEEEauthorrefmark{1}, Le Yu\IEEEauthorrefmark{2}, Sen Chen\IEEEauthorrefmark{3}, Hao Zhou\IEEEauthorrefmark{2}, Xiapu Luo\IEEEauthorrefmark{2}, Shuyue Li\IEEEauthorrefmark{1}, Yang Liu\IEEEauthorrefmark{3}, Jun Liu\IEEEauthorrefmark{1}, Ting Liu\IEEEauthorrefmark{1}
	\\
	\IEEEauthorblockA{\IEEEauthorrefmark{1}School of Cyber Science and Engineering, MoE KLINNS, Xi'an Jiaotong University, China}
	\IEEEauthorblockA{\IEEEauthorrefmark{2}Department of Computing, The Hong Kong Polytechnic University, China}
	\IEEEauthorblockA{\IEEEauthorrefmark{3}School of Computer Science and Engineering, Nanyang Technological University, Singapore}
	\IEEEauthorblockA{mingfan@mail.xjtu.edu.cn; yulele08@gmail.com; ecnuchensen@gmail.com; cshaoz@comp.polyu.edu.hk;\\ luoxiapu@gmail.com; lishuyue1221@stu.xjtu.edu.cn; yangliu@ntu.edu.sg; liukeen@xjtu.edu.cn; tingliu@mail.xjtu.edu.cn
		}
}

\maketitle

\begin{abstract}
The purpose of the General Data Protection Regulation (GDPR) is to provide improved privacy protection. If an app controls personal data from users, it needs to be compliant with GDPR. However, GDPR lists general rules rather than exact step-by-step guidelines about how to develop an app that fulfills the requirements.
Therefore, there may exist GDPR compliance violations in existing apps, which would pose severe privacy threats to app users.
In this paper, we take mobile health applications (mHealth apps) as a peephole to examine the status quo of GDPR compliance in Android apps.
We first propose an automated system, named \mytool, to bridge the semantic gap between the general rules of GDPR and the app implementations by identifying the data practices declared in the app privacy policy and the data relevant behaviors in the app code. Then, based on \mytool, we detect three kinds of GDPR compliance violations, including the incompleteness of privacy policy, the inconsistency of data collections, and the insecurity of data transmission. We perform an empirical evaluation of 796 mHealth apps. The results reveal that 189 (23.7\%) of them do not provide complete privacy policies. Moreover, 59 apps collect sensitive data through different measures, but 46 (77.9\%) of them contain at least one inconsistent collection behavior.
Even worse, among the 59 apps,  only 8 apps try to ensure the transmission security of collected data.  However, all of them contain at least one encryption or SSL misuse.
Our work exposes severe privacy issues to raise awareness of privacy protection for app users and developers.
\end{abstract}


\section{Introduction}
\label{sec:intro}

The General Data Protection Regulation (GDPR)  is an important data and privacy law, enforced since May 2018. 
The purpose of GDPR is to provide improved privacy protection based on a set of standardized data protection laws. For mobile apps, the GDPR applies to ones that collect and process personal data of European Union (EU) citizens. It does not matter if the app is operated from outside of the EU. The GDPR will still apply. Therefore, the GDPR is of considerable significance to mobile apps.

However, the regulation in GDPR itself does not contain any exact step-by-step guidelines about how to develop an app that fulfills all the requirements. It only gives us a list of the general rules that we must keep in mind when creating apps. Therefore, the semantic gap between the general rules and the app implementations may result in compliance violations between GDPR and apps, which would pose severe privacy threats to app users. Once that happens, the developers would face administrative fines of up to to 20 million EUR, or in the case of an undertaking, up to 4\% of the total worldwide annual turnover of the preceding financial year, whichever is higher~\cite{GDPRFine}.


In this paper, we take mHealth apps as a peephole to investigate  the status quo of GDPR compliance in the Android apps. The rationale is that the mHealth apps, which are developed to perform health-related activities to help users monitor and manage their state of health, usually collect a broad range of more critical health-related information (PHI)~\cite{PHIDic} compared with the apps of other categories. Moreover, PHI is an important special category of personal data protected by GDPR. 

We carefully read the articles in GDPR and summarize the following three necessary regulations based on three basic requirements for data protection. (see details in Section \ref{subsec:gdpr}).

\noindent$\bullet$ {\textbf{Completeness of Privacy Policy.} The app should provide a  complete privacy policy~\cite{AndroidPrivacy} to inform users about how their data is collected and used before app installation.}

\noindent$\bullet$ {\textbf{Consistency of  Data Collection.} The app should not access more data than what its privacy policy declares.}

\noindent$\bullet$  {\textbf{Security of Data Transmission.} The data transmission of an app should adopt reasonable measures to keep secure.}

%
%

However,  it is difficult to investigate compliance violations with the three regulations due to three challenges.

First, for the detection of the incomplete privacy policy, since the policies are written in natural language without a uniform structure, it is difficult to understand the relations between the semantics declared in the privacy policy and the notices declared in GDPR.
Most of the existing approaches only focus on the analysis of privacy policy with the natural language processing (NLP) techniques \cite{xiao2012automated, costante2013websites, brodie2006empirical} without considering the GDPR.
To address this challenge, we combine the NLP techniques with the machine learning algorithms to generate six notice classifiers to detect whether a privacy policy is complete or not (see Section~\ref{subsec:incomplete} for details).

Second, for the analysis of the inconsistent data collection, the PHI of mHealth apps is usually entered by users through the graphical user interface (GUI), making the related studies that only concentrate on the system-managed user data (e.g., device id, IMEI, and IP address) obtained by API calls inapplicable~\cite{yu2016can, slavin2016toward, zimmeck2017automated}.
To address this challenge, we analyze the app GUI to recognize the PHI inputs and record its semantics based on a predefined PHI keyword database.
Then, we leverage the static analysis technique to track the data flow of PHI to identify whether the mHealth app collects it by writing to files or sending out.
Finally, we match the collected PHI with the those declared in privacy policy to identify whether there exists an inconsistent behavior of data collection (see Section \ref{subsec:inconsistent} for details).

Third, for the identification of the data transmission security, it is challenging to map abstract cryptographic concepts (i.e., standard security rules) to concrete program properties.
Existing approaches~\cite{egele2013empirical, fahl2012eve, nguyen2017stitch} can only check violations of security rules regardless of what data is protected;
thus, they cannot be directly adopted by our work.
Therefore, we initially analyze the tracked data flow information and identify which PHI is protected by cryptographic implementations.
Then we study whether the implementations are compliant with standard security rules (see Section \ref{subsec:insecure} for details).

We implement the above ideas in a new system called \mytool{} to detect the GDPR compliance violations in mHealth apps.  We conduct an empirical study on 796 real mHealth apps to examine whether they are compliant the regulations. The main contributions are as follows:





\renewcommand\theenumi{\roman{enumi}}
\renewcommand\labelenumi{(\theenumi)}

\begin{enumerate}
	\item {To our best knowledge, this is the first systematic investigation on automatically detecting the compliance violations between the GDPR and mHealth apps.}
	\item{We propose and develop \mytool{}, an automated system to effectively detect whether mHealth apps are compliant with three  privacy regulations summarized from GDPR.}
	\item{We conduct an empirical evaluation with \mytool{} on 796 real mHealth apps. The experimental results show that \mytool{} can effectively detect the regulation violations for mHealth apps, which has exposed severe privacy issues to raise the awareness of  privacy protection for the mHealth app users and developers. }
\end{enumerate}


\section{Background and Problem Definition}\label{sec:back}

\subsection{GDPR}
\label{subsec:gdpr}
The GDPR agreed upon by the European Parliament and Council in April 2016, has replaced the Data Protection Directive 95/46/ec in May 2018 as the primary law~\cite{GDPR}. It consists of 11 chapters and 99 articles that regulate how organizations collect, use, share, secure, and process their personal data and privacy of EU citizens for transactions that occur within EU member states.  Organizations that fail to achieve GDPR compliance before the deadline (i.e., May 2018) will be subject to stiff penalties and fines. Note that even organizations outside the EU need to be compliant if they offer services to EU citizens. Never before have the needs of app users in this area been so forcefully and comprehensively protected. The seriousness of the GDPR should not be underestimated, which is why we have to take a fresh look at the  compliance problem between the GDPR and Android apps. 
The analysis scope of this work focuses on three regulations
that are summarized from three corresponding basic requirements in GDPR Article 5, i.e., transparency, data minimization, and confidentiality. 

Transparency: the GDPR requires that \textit{all the information you provide about your data processing needs to be easy to access and easy to understand}. Providing a privacy policy is an effective and necessary way to improve transparency~\cite{Inbenda}. Thus, we analyze the data processing transparency by detecting the completeness of privacy policy. 

Data minimization: the GDPR requires that \textit{personal data shall be adequate, relevant and limited to what is necessary in relation to the purposes for which they are processed.} Thus, we analyze the consistency of data collection by detecting whether the app has accessed more data than what its privacy policy declares.

Confidentiality: the GDPR requires that \textit{personal data shall be processed in a manner that ensures appropriate security of the personal data.} Thus, we check  whether the data transmission of an app is adopted reasonable measures.  

Note that there are many requirements defined in GDPR, and it is challenging to consider all the requirements. In this work, we only check  
	three fundamental regulations based on the above three basic requirements for data protection. For other regulations, we leave them as our future work.


%
%

\subsection{Privacy Policy}
\label{subsec:PrivacyPolicy}
\begin{figure}[t]
	\centering
	\includegraphics[width=0.95\linewidth]{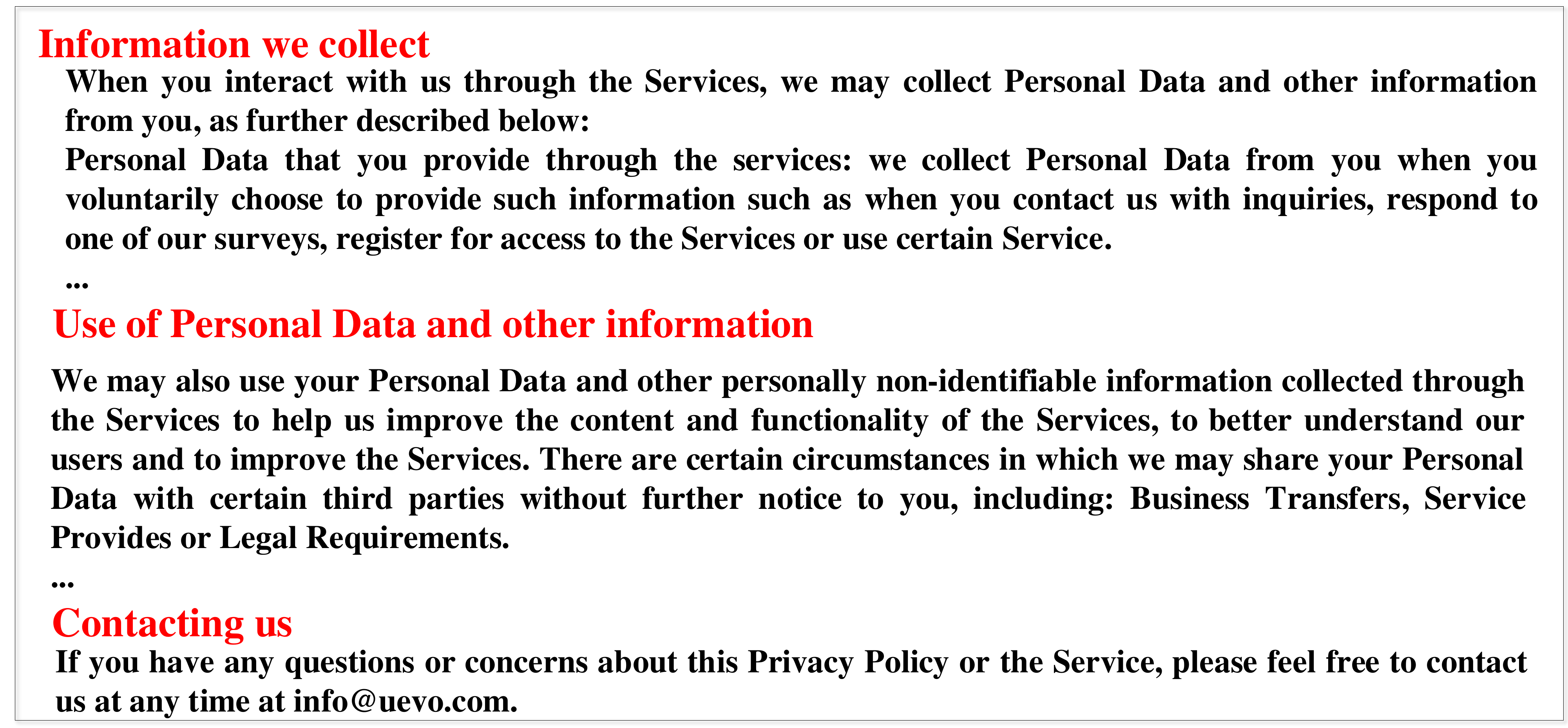}
	\caption{A privacy policy of app called \textit{com.uevo.heartrate}}
	\label{Fig-Sec2-PrivacyPolicy}
	\vspace{-20pt}
\end{figure}

Software that operates on personal data is often required to be accompanied by a \emph{privacy policy}, which is a legal document and software artifact that describes consumer data practices~\cite{bhatia2016mining}.
For Android platform, the privacy policy is designed and uploaded to {Google Play Store}~\cite{GooglePlay} by relevant developers for declaring what user data will be collected, why it will be collected, and how it will be used~\cite{AndroidPrivacy}.
Fig.~\ref{Fig-Sec2-PrivacyPolicy}  presents the privacy policy of an app called \textit{com.uevo.heartrate}.
The privacy policy initially declares that the app will collect personal data from the users when they voluntarily choose to provide such data.
Then, it describes that the app developers may share personal data with certain third parties without further notice to users.
Finally, the developers provide their contact information. Note that the policies are generally ambiguous, resulting a challenge to directly understand the semantic meanings~\cite{bhatia2016a,evans2017an}.  

\subsection{Research Questions}
\label{subsec:problem}

\noindent\textbf{RQ 1: Do the mHealth apps provide complete privacy policies?}

To improve the transparency of data processing,  the mHealth app developers need to provide a privacy policy that at least:

\begin{itemize}
	\item {Indicates the precise categories of personal data that the app will process (\textit{\textbf{Data Collection}});}
	\item {Describes the purpose of data processing, and how the data will be used and fitted in the products and services (\textit{\textbf{Data Usage}});}
	\item {Informs the user of their right to access and correct personal data, and to delete it (\textit{\textbf{User Right}});}
	\item {Informs the user that their use of the app is strictly voluntary, but requires their consent to permit the processing of personal data (\textit{\textbf{User Consent}});}
	\item {Informs that appropriate technical measures are adopted to protect personal data (\textit{\textbf{Data Security}});}
	\item {Provides contact information where the user can ask data protection related questions (\textit{\textbf{Contact Information}}).}
\end{itemize}

Note that even the above six notices provided here are not complete compared with the work~\cite{torre2020ai}; they are the minimal fundamental requirements that a complete policy should contain according to the official GDPR privacy notice template~\cite{PrivacyNotice}.


The first goal  is to automatically detect whether a given privacy policy contains the six minimal notice specifications.
We use $Notice_{pp}$ to denote the set of contained notice category labels of a privacy policy $pp$. Complete privacy policy means its $Notice_{pp}$ contains all the six labels, i.e., $|Notice_{pp}|=6$.
For example, by carefully scrutinizing the privacy policy illustrated in Fig.~\ref{Fig-Sec2-PrivacyPolicy}, we observe that it contains only three notice specifications and its $Notice_{pp}=\{$\textit{Data Collection, Data Usage, Contact Information}$\}$. Therefore, the app does not provide a complete privacy policy.
To achieve the goal, the main limitation is that the privacy policies are not written in a structured, commonly used, and machine-readable format, resulting that we cannot directly obtain $Notice_{pp}$.


\noindent\textbf{RQ 2: Do the mHealth apps declare all the collected PHI in their privacy policies?  }

Every mHealth app must be designed to only collect and process PHI for its specific and legitimate purpose,  which are clearly defined in the privacy policy.
Here we define the PHI is collected by a mHealth app if it is input by users, and stored with different channels such as sending out through network or writing in local files.
Collecting more PHI than what it declares is an illegal behavior to the app users.
Furthermore, the data breach of such collected PHI could seriously affect the app users' profits and their confidence in the mHealth app.

The second goal is to automatically discover whether there is any PHI collected by a mHealth app but not declared in the privacy policy.  To this end, we need to obtain two PHI sets. One is the set of declared collected PHI ($DCP$) extracted from the privacy policy. The other is the set of actually collected PHI ($ACP$) discovered from the app code. Thus, one inconsistent behavior is discovered if there is any item in $ACP$ that is not contained in $DCP$.

\begin{figure}[!t]
	\centering
	\includegraphics[width=0.5\linewidth]{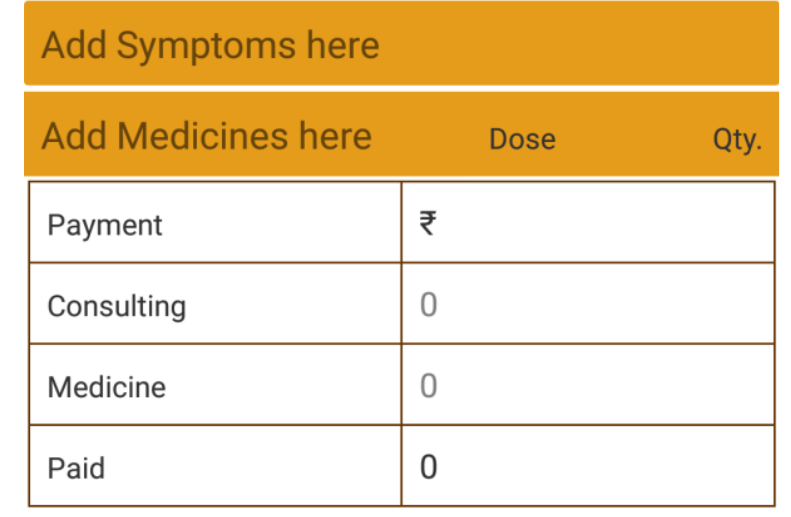}
	\caption{GUI of app called \textit{com.sattva.sattvamanager}} 
	\label{Fig-Sec2-GUIExample}
	\vspace{-20pt}
\end{figure}

The $DCP$ can be constructed by analyzing the PHI items declared in the privacy policy with NLP techniques. However, it is challenging to construct the $ACP$ because the PHI is generally input by users on app GUI~\cite{yu2016can, slavin2016toward, zimmeck2017automated}. For example, Fig. \ref{Fig-Sec2-GUIExample} presents the GUI of an app called \textit{com.sattva.sattvamanager} that requires the users to provide the PHI such as symptoms, payment, and medicine. It is difficult for machines to automatically recognize the content of what user input due to the lack of fixed structures of the GUI.

\begin{figure*}
	\centering
	\includegraphics[width=0.7\linewidth]{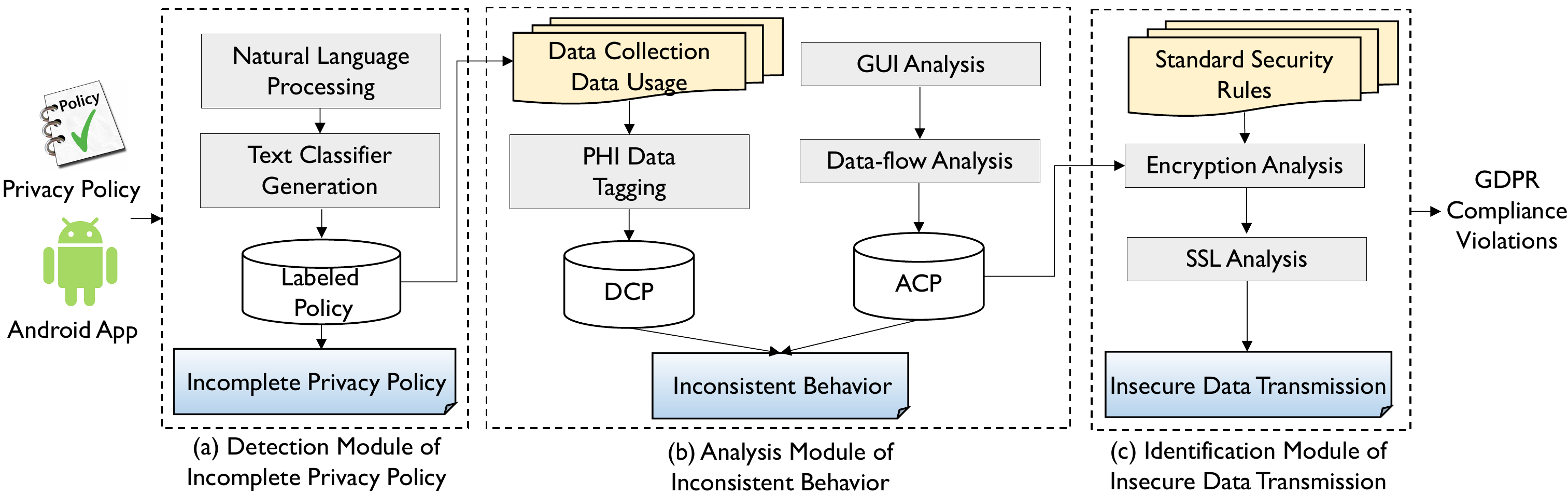}
	\caption{Architecture of \mytool{}.}
	\label{Fig-Sec3-Overview}
	\vspace{-20pt}
\end{figure*}

\noindent\textbf{RQ 3: Do the mHealth apps implement reasonable measures to ensure the transmission security of  their collected PHI?}

The GDPR requires the developers to provide appropriate technical safeguards to ensure the transmission security of collected PHI. Encryption is the most obvious determinant of security in mHealth apps communications \cite{Encryption}. 
However, even for the encrypted PHI, there might be encryption misuse that is not compliant with standard security rules. 

Here, we make a list of the most common encryption misuses according to the existing studies~\cite{ECB, Hash, IV, nguyen2017stitch, random}:
\ding{172}  Do not use electronic codebook (ECB) mode for encryption \cite{ECB}; ECB mode uses a weak encryption algorithm that produces the same ciphertext from the same plaintext blocks, which would allow attackers to  gain the sensitive data easily.
\ding{173}  Do not use MD5 or SHA-1 algorithms for encryption \cite{Hash}; The modern attacks can compute large numbers of hashes, or even exhaust the entire space of all possible passwords using massively-parallel computing.
\ding{174} Do not use a constant  initialization vector (IV) \cite{IV} or constant keys \cite{nguyen2017stitch} for encryption; The constant IV or keys result in a deterministic and stateless cipher, which would make the information insecure.
\ding{175}  Do not use the static seed for \textit{SecureRandom()} API call while generating random number \cite{random}; Using the static seed is predictable and can result in the generation of random numbers with insufficient entropy.

Another common approach to protecting data during communication on the Android platform is to use the Secure Socket Layer (SSL) or Transport Layer Security (TLS) protocols. For brevity, we refer to both protocols as SSL.  The inadequate use of SSL can be exploited to launch Man-in-the-Middle attacks \cite{jackson2008protecting, song2010peeping}.

Note that the above encryption misuses are not specified in GDPR, but they are essential requirements that we need to satisfy to ensure transmission security, which is an important concept in GDPR.

The third goal  is to identify whether there is unprotected PHI, and misuse of the technical safeguards. The main challenge is the mapping of abstract cryptographic concepts introduced above to concrete program properties. Even existing approaches \cite{egele2013empirical, fahl2012eve, nguyen2017stitch} have conducted  related studies; they cannot be directly adopted here since they only check violations of security rules regardless of what data is protected.

\section{Methodology}\label{sec:methodology}
The overview architecture of \mytool{} is illustrated in Fig. \ref{Fig-Sec3-Overview}, which  consists of three main modules.

\subsection{Detecting Incomplete Privacy Policy}
\label{subsec:incomplete}

\noindent{\bf Natural Language Processing.}
Given a privacy policy crawled from the websites in HTML format, we use \textsc{jsoup}~\cite{Jsoup}, a Java HTML parser, to extract the content from the HTML file, and remove all non-ASCII symbols.
Then we split the extracted content into a set of sentences using \textsc{Stanford typed dependency parser}~\cite{StanfordParser}.
After that, each sentence is formalized using the bag-of-words model~\cite{zhang2010understanding} with the word stemming and removing of stop words.
Finally, a sentence $st$ is represented as a feature vector, where each dimension corresponds to the occurrence of a separate word, and the number of dimensions denotes the total number of unique words extracted from privacy policy corpus. If a word occurs in the sentence, its value in the vector is 1; otherwise, the value is 0.

\noindent{\bf Machine Learning Classification.}
After the prepossessing of the privacy policy, we leverage the machine learning algorithms to predict what notices do the extracted sentences belong to.
We use the term $Notice_{st}$ to denote the set of notice labels for sentence $st$.
To generate the classifiers used for notice prediction, we first need to  construct a ground truth dataset manually.
In detail, we initially select 100 privacy policies of mHealth apps that are crawled from the Google Play Store and extract all corresponding sentences.
Then, two co-authors go through these sentences and understand the semantics of each sentence, and manually construct their corresponding $Notice_{st}$. Note that, for each sentence, if the constructed $Notice_{st}$ by the two  co-authors  are not same, then a third co-author will check the result by having a discussion with them to  guarantee the labeling correctness. In this way, we are able  to 
 obtain a ground truth dataset of sentences that are attached with notice labels. The description of the dataset is listed in Table \ref{Tab-Sec4-classifyExample}.
Each notice category contains at least 100 sentences, and there are 1,284 labeled sentences in total. Note that the construction of labeled ground truth with manual intervention is once for all when training the classifiers. For the other modules, we do not need additional manual intervention.

\begin{table}[t]
	\centering
	\scriptsize
	\caption{Descriptions of the constructed ground truth}
\begin{tabular}{lclc}
	\toprule
	\textbf{Notice Category} &  \textbf{\#Sentences} & \textbf{Notice Category} &  \textbf{\#Sentences} \\ \toprule
	\textit{\textbf{Data Collection}} & 385  & \textit{\textbf{User Consent}} & 121 \\
	\textit{\textbf{Data Usage}} & 334 & \textit{\textbf{Data Security}} & 176 \\
	\textit{\textbf{User Right}} & 164 & \textit{\textbf{Contact Information}} & 104\\
	\bottomrule
\end{tabular}
	\label{Tab-Sec4-classifyExample}
	\vspace{-15pt}
\end{table}

Next, we construct a classifier for each notice prediction based on the dataset. For example, to construct the classifier used for \textit{Data Collection} notice  prediction, the 385 sentences that contain the \textit{Data Collection} label are regarded as the positive instances. In addition, equal size of negative instances are randomly selected from other labeled sentences.
After that, with the state-of-the-art machine learning algorithms such as Random Forest~\cite{breiman2001random}, the classifier used to predict the \textit{Data Collection} notice can be generated.

\noindent{\bf Incomplete Policy Detection.}
After the generation of six classifiers, the feature vector of a new sentence will be put into the six classifiers in sequence. If the prediction result of a notice classifier is 1, then its corresponding notice label will be put into the  $Notice_{st}$ of the given sentence. Finally, the $Notice_{pp}$ is obtained by merging all the generated $Notice_{st}$ to detect whether the privacy policy is complete. 


\vspace{-5pt}
\subsection{Analyzing Inconsistent Behavior}
\label{subsec:inconsistent}


Before the construction of $DCP$ and $ACP$, it is essential to construct a set of PHI.
To this end, we carefully read the GDPR recital 54~\cite{Recital54} and the National Health Data Dictionary provided by the Australian Institute of Health and Welfare~\cite{PHI}, and then add the concrete PHI items into a set.
We use $PS=\{ps_i|1\le i \le m\}$ to denote the full set of PHI, where $ps_i$ denotes the unique name of a PHI item in $PS$, and $m$ denotes the number of PHI items; $m=227$ in this work.


\subsubsection{DCP Construction}

To construct $DCP$, we focus on the sentences with notice labels of \textit{Data Collection} and \textit{Data Usage} since we observe that all the data related operations are declared in them.
We first leverage the \textsc{Stanford typed dependency parser} \cite{StanfordParser} to extract all the noun phrases. However, it is not effective to directly map the noun phrases with the items in the $PS$ due to the diversity of natural language. For example, the PHI called ``doctor name" might be written as ``name of doctor'' in the privacy policy.

\noindent{\bf Phrase Similarity Calculation.}
To measure the similarities between the semantics of PHI with the noun phrases extracted from the privacy policy, we rely on the tool called \textsc{Word2Vec}~\cite{mikolov2013distributed} to transform the PHI items and the noun phrases into a calculable form. As a result, the similarities can be obtained based on the 
cosine similarity.  If the similarity is higher than the threshold $\epsilon$, which is set as 0.85, then $ps_i$ is added into the $DCP$. Note that the parameter 0.85 is preset based on our experience analysis on a set of similar phrases.


\subsubsection{ACP Construction}

The construction of $ACP$ relies on two kinds of techniques: the GUI analysis technique, which is used to identify the user input PHI; the data-flow analysis technique, which is used to filter out the non-collected PHI.


\noindent{\bf GUI Analysis.}
To analyze the user input PHI from the GUI of Android app, we need to render all the GUI layouts and identify the semantics of the user inputs. Note that our technique is extended based on UiRef~\cite{andow2017uiref}.
First, an APK file is disassembled using \textsc{apktool}~\cite{apktool}. A file called \textit{public.xml} is generated to store all the resource identifiers of layout widgets.
Subsequently, a customized activity is injected into the APK, and the manifest file of the app is rewritten to register the injected activity as an entry point.
After that, the APK is reassembled and installed on a live device. When the injected activity is launched, the layouts of the app are rendered iteratively by invoking the \textit{setContentView()} API call.
However, the dynamically generated text (e.g., label text set by the \textit{setText()} API call) cannot be extracted in this way using  UiRef~\cite{andow2017uiref}.
To solve the problem, we also launch the activities declared in the \textit{AndroidManifest.xml} file to render the corresponding layouts.
Once a current layout is loaded, its view hierarchy is dumped by \textsc{UIAutomator}~\cite{UIAutomator}.
Then, the associated metadata of UI widgets are extracted from the view hierarchy, and each widget is represented as a four tuples form $\langle class, id, text, bound \rangle$, where

\begin{itemize}
	\item {$class$ denotes the widget type such as \textit{EditText}.}
	\item {$id$ denotes the widget id stored in the \textit{public.xml} file.}
	\item {$text$ denotes the plain text presented in the widget.}
	\item {$bound$ denotes the coordinate of the displayed  widget.}
\end{itemize}

\begin{figure}[t]
	\centering
	\includegraphics[width=\linewidth]{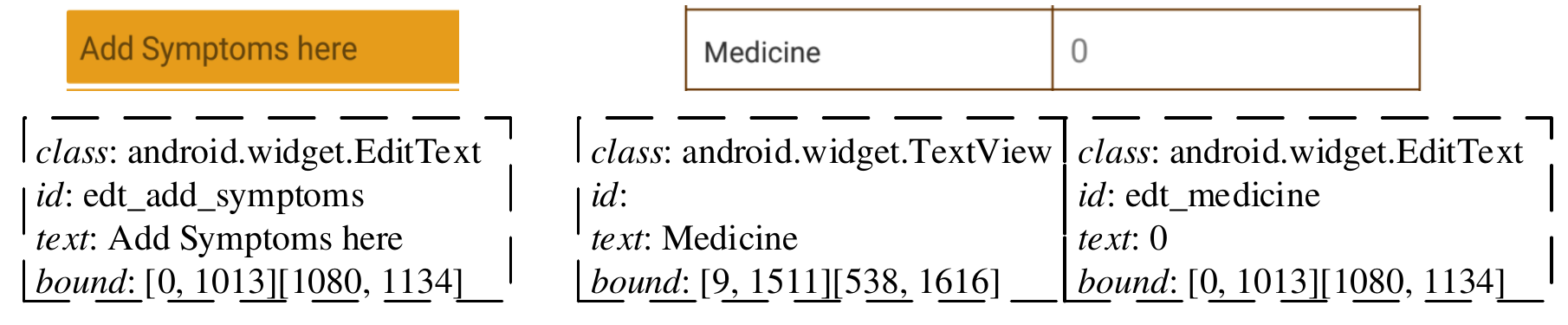}
	\caption{Two examples of user input widgets and their associated metadata. }
	\label{Fig-Sec3-Layout}
	\vspace{-20pt}
\end{figure}


Next, we need to understand the semantics of the inputs. In general, there are two methods for developers to present the semantics of the inputs to help users understand what they are required to provide. The first method is presenting the semantics based on the hint text of the user input widget.  The second method is leveraging a label widget to present the semantics. We call the two methods as the hint-based method and the label-based method.

Two examples are illustrated in Fig. \ref{Fig-Sec3-Layout}. The first example presents an \textit{EditText} that uses the hint text ``Add symptoms here'' belongs to the hint-based method. The second example presents an \textit{EditText} that shows  semantics with a combined \textit{TextView} belongs to the label-based method.

For the hint-based method, we analyze the $text$ value of the user input widget. The string value of the \textit{text} is preprocessed by the NLP technique. Then  it is split into a set of words.  
With such words, we construct a set of unigram phrases and a set of bigram phrases. After that, each phrase in the two sets is matched with the PHI items in $PS$ based on the introduced phrase similarity calculation method. If $ps\in PS$ is matched with a phrase, $ps$ is regarded as one user input PHI. If no $ps$ is matched, the given widget might use the label-based method.

For the label-based method, the main challenge is to map the labels with their combined user input widgets.
In our work, we propose algorithm \ref{labelMap} to map the labels with user input widgets.
Algorithm \ref{labelMap} requires two inputs,  $LB$ and $UIW$. For each input widget in the $UIW$, it is checked whether there are any labels in $LB$ that are closely placed on its left or above by using the function ConstructLeftLabelSet() and ConstructUpLabelSet(). The positional relations between the widgets are calculated based on their extracted $bound$ values. Finally, for each pair in the output $M$, the $text$ value of the label will be matched with the PHI items in the same way as solving the hint-based way to identify the user input PHI. 

\begin{algorithm2e}[t]
	\footnotesize
	\setcounter{AlgoLine}{0}
	\caption{Mapping of Labels and Input Widgets}
	\label{labelMap}
	\DontPrintSemicolon
	\SetCommentSty{mycommfont}
	\KwIn{$LB$: the set of labels displayed in an UI;  $UIW$:  the set of user input widgets displayed in an UI.}
	\KwOut{$M$: the set of output label and input widget mapping pairs.}
	
	\ForEach{$input$ in $UIW$}{
		$ LeftSet \gets$ ConstructLeftLabelSet($input$, $LB$)\;
		\If{$LeftSet$.size()>0}{
			$mapLabel\gets  \mathop{argmin}_{label \in LB}$ {dis($label$, $input$)} \;
			$M$.put($input$, $mapLabel$) \;
			$LB$.remove($mapLabel$) and $UIW$.remove($input$) \;
		}
		\Else{$UpSet \gets$ ConstructUpLabelSet($input$, $Label$)\;
			\If{$UpSet$.size()>0}{
				$mapLabel\gets \mathop{argmin}_{label\in LB}$ {dis($label$, $input$)}\;
				$M$.put($input$, $mapLabel$)\;
				$LB$.remove($mapLabel$) and $UIW$.remove($input$) \;
			}
		}
	}
	\Return{$M$}
\end{algorithm2e}

\noindent{\bf Data-flow Analysis.}
Note that the user input PHI cannot be directly added into the $ACP$ since the app might not store them.
Thus, in our approach, the data-flow analysis technique is used to recognize the point of getting specified data and to further check the destination of fetched data. We conduct the data flow analysis based on static analysis tools such as \textsc{FlowDroid} \cite{arzt2014flowdroid}
and \textsc{VulHunter} \cite{qian2015vulhunter}, which are implemented based on the Soot framework \cite{vallee1999soot}. Notably, to enhance the static analysis accuracy, \textsc{IccTA} \cite{li2015iccta} is employed to identify the source and the target of intents, and \textsc{EdgeMiner} \cite{cao2015edgeminer} is utilized to determine the implicit callbacks (e.g., from \textit{setOnClickListener()} to \textit{onClick()}). The data-flow analysis includes three main parts, listing as below:

\begin{itemize}
	\item{\textbf{Data Sources.} The data sources are the points that we obtain the PHI which will be tracked. We focus on the user input PHI and the API call \textit{findViewById()} is selected as the data source.}
	
	\item{\textbf{Data Propagation.} To track the data propagation in the app code, we leverage the taint propagation techniques \cite{arzt2014flowdroid}. In detail, the data sources are initially assigned with a unique taint tag. Then, the taint tag will be propagated based on the direct data flow propagations according to the intermediate representation extracted by Soot.}
	
	\item{\textbf{Data Sinks.} The data sinks are the data use points of the tainted variables. There are six different kinds of data storage methods, including writing data into a log (e.g., \textit{Log.d()}) or a file (e.g., \textit{FileoutputStream.write()}), or sending data out through network (e.g., \textit{HttpClient.execute()})) or short messages (e.g., \textit{SmsManager.sendTextMessage()}), or inserting the data into a database (e.g., \textit{SQLiteDatabase.update()}) or the content provider (e.g., \textit{ContentResolver.insert()}).}
\end{itemize}

Note that, we also need to link the data flows with their corresponding widget objects.
We resolve the argument value of the \textit{findViewById()} API call, and the argument value demonts the \textit{id} of the specific wedget object.

In summary, if an app collects one PHI and stores the data with the above six methods, the PHI is added into $ACP$.
Finally, one inconsistent behavior is discovered if there exists an item in $ACP$ that is not contained in $DCP$.

\subsection{Identifying Insecure Data Transmission}
\label{subsec:insecure}

\subsubsection{Encryption Analysis}
For the identification of the use of system-provided encryption algorithms, we observe that the symmetric encryption (e.g., AES algorithm) and asymmetric  encryption (e.g., RSA algorithm) schemes are generally accessible to an app through the \textit{Cipher} object by using the \textit{doFinal()} API call. In addition, the one-way encryption schemes (e.g., MD5 and SHA-1 algorithms) are generally accessible to an app through the \textit{MessageDigest} object by using the \textit{digest()} API call. Therefore, we use such encryption API calls as data sinks and apply data-flow analysis techniques to detect whether there exists complete data flow from the data sources (i.e., \textit{findViewById()}) to any encryption API calls. If no complete data flow is found, we regard that such PHI is not protected with system-provided encryption algorithms.

Although there are security pieces of advice in the official documents and an extensive body of security-related research about exploits and vulnerabilities, using encryption algorithm correctly is still not easy for inexperienced or distracted developers~\cite{rahaman2018cryptoguard}. We assess the  four security rules (introduced in Section \ref{subsec:problem}) on the mHealth apps by checking their corresponding program properties.

To evaluate the rule \ding{172}, we resolve the \textit{Cipher.getInstance()} API call to find what transformation string is specified by the developers to be used as arguments of the API call. If the string ``ASE'' is used as the argument, the encryption mode is automatically chosen as ECB mode. To improve security, another encryption mode with padding such as ``AES/CBC/PKCS5Padding'' should be used.

To evaluate the rule \ding{173},
we initially check whether the \textit{digest()} API call is used as the data sink.
If so, we then resolve the \textit{MessageDigest.getInstance()} API call to find whether its argument is specified with string ``MD5'' or ``SHA-1''.

To evaluate the rule \ding{174},
we compute the backward slices for the arguments of \textit{IvParameterSpec()} and \textit{SecreKeySpec()} API calls, and then determine whether the used arguments consist of constant values. If the slices only depend on constant values, the IV or the keys are constant too.

To evaluate the rule \ding{175},
we construct a backward slice from each call site to the \textit{SecureRandom()} API to check whether the developers specify the seed value.

\subsubsection{SSL Analysis}
SSL is another common approach to protect data during transmission on the Android platform, in which the \textit{java.net}, \textit{android.net} and \textit{org.apache.http} packages can be used to create sockets or HTTP(S) connection. However, as introduced in \cite{aha1991instance}, a large number of apps implement SSL with inadequate validation such as containing code that allows all hostnames or accepts all certificates. Insecure SSL transmission is dangerous since they would generally carry critical sensitive data  such as the collected PHI. To detect the usage of SSL analysis in mHealth apps, we focus on the PHI that is sent out through the network with sink API calls in the three network-related packages. Then we identify the SSL misuse by using a static analysis tool called \textsc{MalloDroid} \cite{fahl2012eve}, which can automatically check SSL security risks in apps.

\section{Evaluation}
\label{sec:evaluation}


\subsection{Data Collection}

To evaluate \mytool{} we initially crawl 1,200 popular real mHealth apps from \textit{Medical} and \textit{Health} categories on the \textit{Google Play} \cite{GooglePlay} according to their download counts.
Then we remove the apps that do not contain a privacy policy written in English language.
Finally, there are 796 mHealth apps remained.
After that, to answer \textbf{RQ 1}, we use the constructed ground truth dataset (i.e., 1,284 labeled sentences in total) introduced in Section \ref{subsec:incomplete} to generate notice classifiers, and then use the classifiers to detect whether the unlabeled policies are complete or not. To answer \textbf{RQ 2}, we focus on the apps that require users to input PHI on the GUI. After the removing of the apps that are only used to introduce health-related knowledge based on the GUI analysis step, 253 remaining mHealth apps (31.78\% of 796 apps) are  analyzed in the analysis module of inconsistent behavior. To answer \textbf{RQ 3}, the 59 apps that collect PHI are put in the identification module of insecure data transmission to check whether they have implemented reasonable PHI data protection measures.

In addition to the mHealth apps, we also need to construct the set of sink API calls. Based on the API list provided by \textsc{SuSi} \cite{rasthofer2014machine}, we manually remove the API calls that do not belong to our six defined data storage methods. In the end, 78 sink API calls are used.

The apps and their privacy policies, as well as the generated data flow information, can be found on our website\footnote{\url{https://drive.google.com/drive/folders/18qaSTuHcm_2LLBsMM70Y9VwZrfvi686t?usp=sharing}}.

\subsection{RQ 1: Do the mHealth apps provide complete privacy policies?}
\label{subsec:evaNotice}


\begin{table}[t]
	\centering
	\scriptsize
\caption{Classification results for the notice categories in labeled dataset.}
\begin{tabular}{ccccccc}
\toprule[1.5pt]
\multicolumn{1}{c}{\textbf{{\begin{tabular}[c]{@{}c@{}}Notice \\ Category\end{tabular}}}} & \multicolumn{1}{l}{\textbf{Classifier}} & \multicolumn{1}{l}{\textbf{TPR}} & \multicolumn{1}{l}{\textbf{FPR}} & \multicolumn{1}{l}{\textbf{Precision}} & \multicolumn{1}{l}{\textbf{Recall}} & \multicolumn{1}{l}{\textbf{F-1}} \\ \midrule[1pt]
\multirow{3}{*}{\textit{\textbf{\begin{tabular}[c]{@{}c@{}}Data \\ Collection\end{tabular}}}} & \textbf{RF} & \textbf{0.937} & \textbf{0.132} & \textbf{0.877} & \textbf{0.937} & \textbf{0.906} \\
 & NB & 0.882 & 0.153 & 0.852 & 0.882 & 0.867 \\
 & DT & 0.789 & 0.192 & 0.804 & 0.789 & 0.797 \\ \hline
\multirow{3}{*}{\textit{\textbf{\begin{tabular}[c]{@{}c@{}}Data \\ Usage\end{tabular}}}} & \textbf{RF} & \textbf{0.884} & \textbf{0.154} & \textbf{0.852} & \textbf{0.884} & \textbf{0.867} \\
 & NB & 0.865 & 0.164 & 0.841 & 0.865 & 0.853 \\
 & DT & 0.830 & 0.214 & 0.795 & 0.830 & 0.812 \\ \hline
\multirow{3}{*}{\textit{\textbf{\begin{tabular}[c]{@{}c@{}}User \\ Right\end{tabular}}}} & \textbf{RF} & \textbf{0.950} & 0.157 & 0.858 & \textbf{0.950} & \textbf{0.901} \\
 & NB & 0.862 & \textbf{0.113} & \textbf{0.884} & 0.862 & 0.873 \\
 & DT & 0.836 & 0.176 & 0.826 & 0.836 & 0.831 \\ \hline
\multirow{3}{*}{\textit{\textbf{\begin{tabular}[c]{@{}c@{}}User \\ Consent\end{tabular}}}} & RF & 0.878 & 0.104 & 0.902 & 0.878 & 0.890 \\
 & NB & 0.887 & 0.130 & 0.872 & 0.887 & 0.879 \\
 & \textbf{DT} & \textbf{0.922} & \textbf{0.035} & \textbf{0.964} & \textbf{0.922} & \textbf{0.942} \\ \hline
\multirow{3}{*}{\textit{\textbf{\begin{tabular}[c]{@{}c@{}}Data \\ Security\end{tabular}}}} & \textbf{RF} & \textbf{0.914} & \textbf{0.018} & \textbf{0.980} & \textbf{0.914} & \textbf{0.946} \\
 & NB & 0.908 & 0.043 & 0.955 & 0.908 & 0.931 \\
 & DT & 0.896 & 0.055 & 0.842 & 0.896 & 0.918 \\ \hline
\multirow{3}{*}{\textit{\textbf{\begin{tabular}[c]{@{}c@{}}Contact \\ Information\end{tabular}}}} & RF & 0.980 & 0.020 & 0.980 & 0.980 & 0.980 \\
 & NB & 0.980 & \textbf{0.010} & \textbf{0.990} & 0.980 & 0.985 \\
 & \textbf{DT} & \textbf{0.990} & \textbf{0.010} & \textbf{0.990} & \textbf{0.990} & \textbf{0.990} \\ \bottomrule[1.5pt]
\end{tabular}
\label{Tab-Sec4-trainClassifier}
\vspace{-15pt}
\end{table}

\subsubsection{Performance on Labeled Dataset} A ground truth dataset that consists of 1,284 labeled sentences is used to construct six notice classifiers. To select the proper classifier with best detection performance, we apply three widely-used classification algorithms~\cite{zimmeck2017automated} (i.e., Decision Tree (DT) \cite{quinlan2014c4}, Random Forest (RF) \cite{breiman2001random} and Naive Bayes (NB) \cite{john1995estimating}) on each notice category with 10-fold cross-validation.  Table~\ref{Tab-Sec4-trainClassifier} lists the results. Using F-1 value to measure the classifier performance, Random Forest performs best in four notice categories while Decision Tree performs best in the remaining two notice categories.
Consequently, the Decision Tree classifier is used to detect the \textit{User Consent} and \textit{Contact Information} categories, while the Random Forest classifier is used to detect the others.


\begin{figure}[t]
	\includegraphics[width=0.42\linewidth]{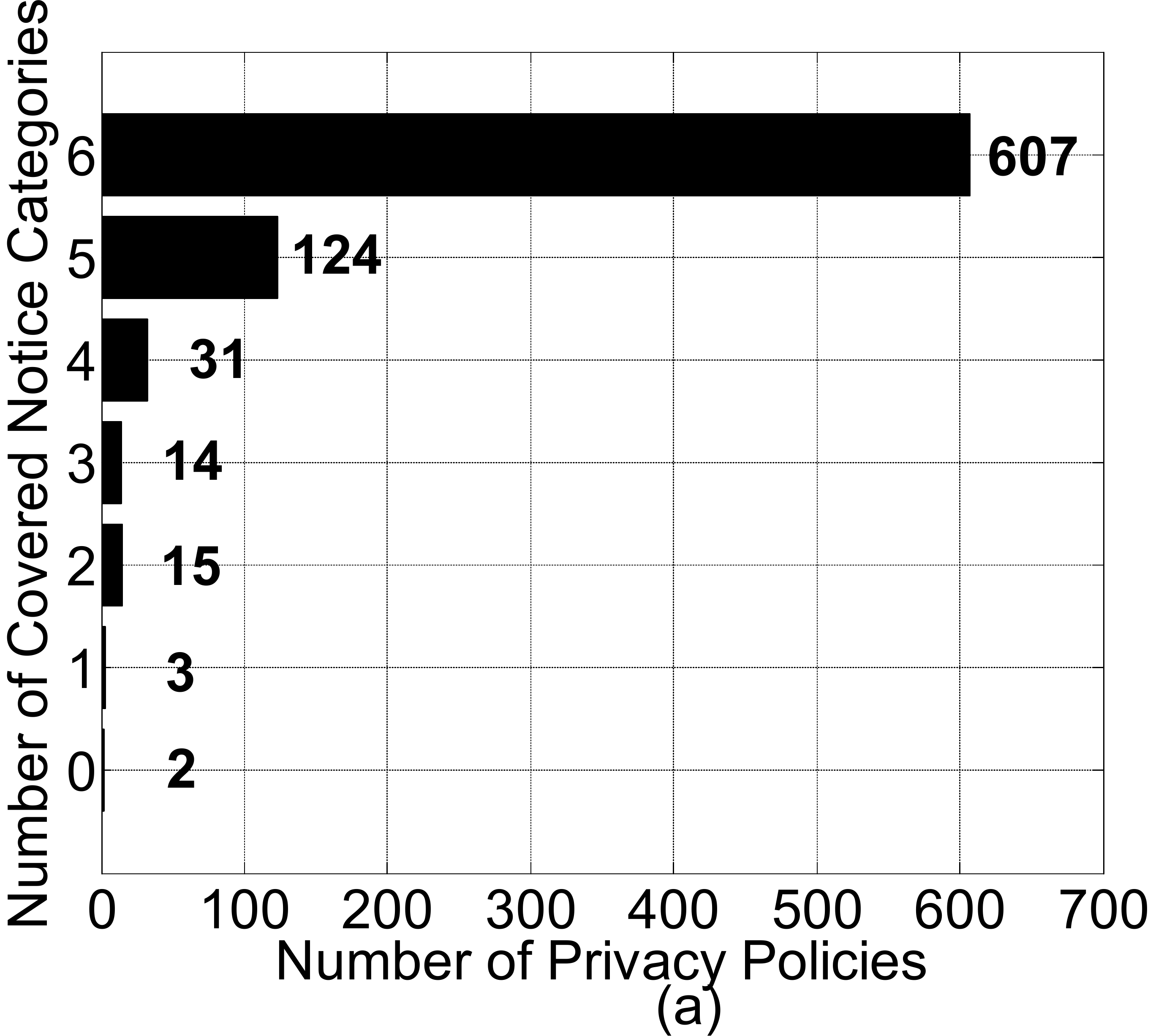}
	\includegraphics[width=0.57\linewidth]{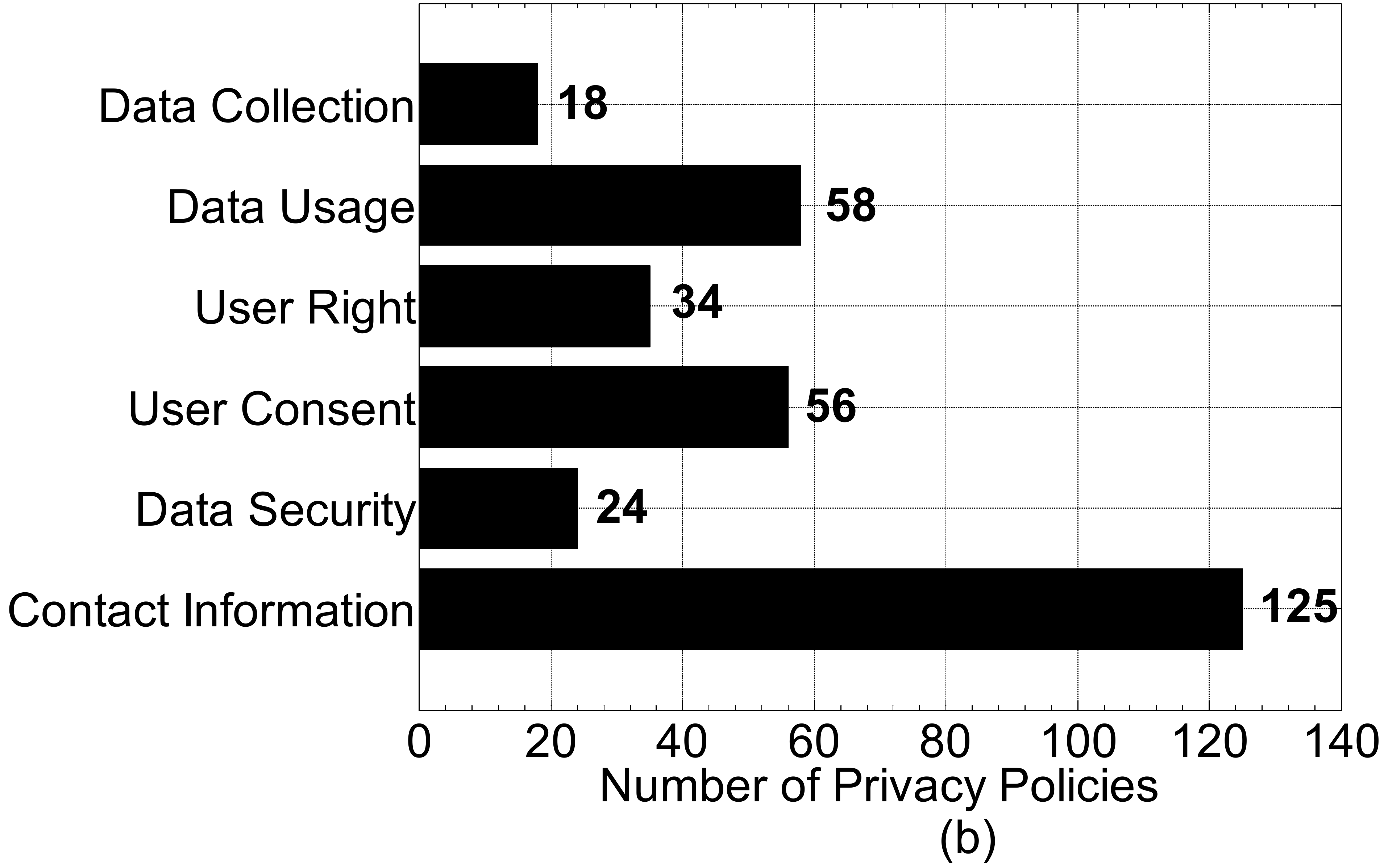}
	\caption{Completeness detection results for 796 unlabeled privacy policies: (a) the number of privacy policies that cover different notice categories; (b) the number of privacy policies that lack of different notice categories.}
	\label{Fig-Sec4-CompletenessDetection}
	\vspace{-20pt}
\end{figure}

\subsubsection{Detection Results on Unlabeled Dataset}
We apply the constructed classifiers on the 796 unlabeled privacy policies to detect whether they are complete.
As illustrated in Fig. \ref{Fig-Sec4-CompletenessDetection} (a), 607 (76.3\%) privacy policies cover all the notice categories and the other 189 (23.7\%) privacy policies are incomplete, of which 34 policies cover no more than three notice categories.
Notably, there are two privacy policies (i.e., \textit{com.bytewaremobile.oasi} and \textit{com.app.tctnews}) that do not cover any notice category, indicating that they do not provide any useful information for  app users. We manually inspect the privacy policy links provided by the two apps, and find that they redirect to  other websites that do not contain any privacy policy related content.
Fig. \ref{Fig-Sec4-CompletenessDetection}(b) presents the number of privacy policies that lack different notice categories. For example, 58 privacy policies that do not cover \textit{Data Usage} notice. More interesting, we find that 48 of the 58 policies have the \textit{Data Collection} notice, indicating that such privacy policies collect personal information but do not  illustrate what purpose the information is used for, which violates the transparency requirement of GDPR. In addition, 125 (15.7\%) privacy notices do not cover the \textit{Contact Information} notice category, which indicates the users could not contact with the app developers if they have any problems about the app usages. The main reason might be that the developers have provided emails on the app downloading page; thus, they think it is not necessary to provide again in the privacy policy.

\begin{figure}[t]
	\centering
	\includegraphics[width=0.85\linewidth]{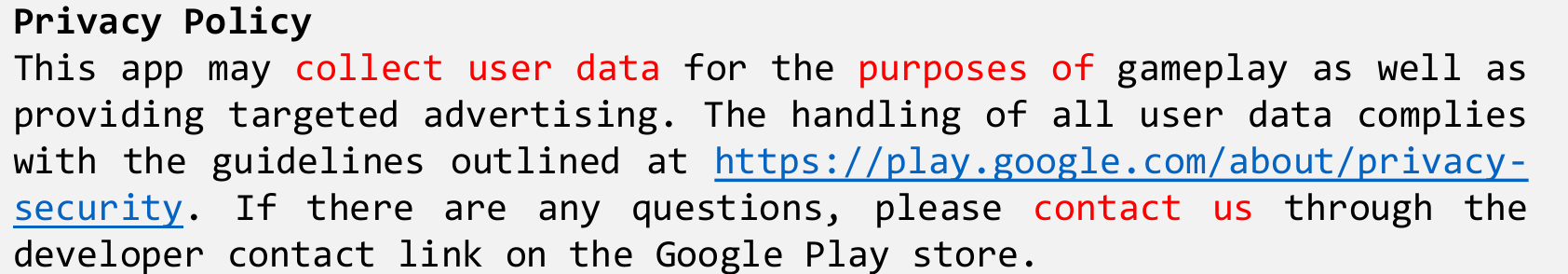}
	\caption{An example of the incomplete privacy policy of app \textit{com.bodyyouwant}}
	\label{Fig-Sec4-Incomplete}
	\vspace{-20pt}
\end{figure}

\subsubsection{False Positives and False Negatives}
We further investigate the false positives and false negatives of the constructed classifiers. To this end, we randomly select 100 sentences from the unlabeled dataset. Then we carefully read the sentences and check whether their attached labels are correct. We find that 7 sentences are incorrectly classified due to the features with high information gain (e.g., ``we'', ``please'', and ``collect'') occur in more than two categories. For example, the sentence \textit{``By using  our  website,  you  agree  that  we  can  place  cookies  on  your  computer/device.''} is attached to labels \textit{User Consent} and \textit{User Right}. The label \textit{User Right} is incorrectly attached since the sentence contains the keywords ``you'' and ``can'' similar to the structures of training sentences in \textit{User Right} category.

Moreover, four sentences are found as false negatives because their verbs do not occur in our labeled training dataset. For example, the verb ``forward'' in the sentence \textit{``The data have never been and will not be forwarded to third parties.''} is not found in the training dataset. To reduce this threat, replacing the uncommon verbs with the common ones through sentence semantic analysis is a promising way.

\begin{table*}[t]
	\centering
	\footnotesize
	\caption{Distribution of the collected PHI for 59 mHealth apps that actually collect PHI. The numbers of the non-declared PHI are listed in the brackets}
	\scalebox{0.9}{
\begin{tabular}{p{1.2cm}<{\centering}p{0.4cm}<{\centering}p{0.4cm}<{\centering}p{0.9cm}<{\centering}p{0.4cm}
		<{\centering}p{0.9cm}<{\centering}p{0.7cm}<{\centering}p{0.7cm}<{\centering}
		p{1.2cm}<{\centering}p{0.4cm}<{\centering}p{0.4cm}<{\centering}p{0.9cm}<{\centering}p{0.4cm}
		<{\centering}p{0.9cm}<{\centering}p{0.7cm}<{\centering}p{0.7cm}<{\centering}
	}
\toprule[1.5pt]
\textbf{PHI} & \textit{\textbf{Log}} & \textit{\textbf{File}} & \textit{\textbf{Network}} & \textit{\textbf{SMS}} & \textit{\textbf{Database}} & \textit{\textbf{Content}} & \textbf{Total} & \textbf{PHI} & \textit{\textbf{Log}} & \textit{\textbf{File}} & \textit{\textbf{Network}} & \textit{\textbf{SMS}} & \textit{\textbf{Database}} & \textit{\textbf{Content}} & \textbf{Total} \\ \midrule[1pt]
name & 23 & 7 & 7 & 0 & 3 & 1 & 41(17) & symptom & 1 & 0 & 0 & 0 & 0 & 0 & 1(1) \\
email & 17 & 1 & 9 & 0 & 0 & 0 & 27(7) & amount & 1 & 0 & 0 & 0 & 0 & 0 & 1(0) \\
password & 13 & 4 & 4 & 0 & 0 & 0 & 21(16) & patient & 0 & 1 & 0 & 0 & 0 & 0 & 1(1) \\
phone & 6 & 1 & 5 & 1 & 0 & 0 & 13(6) & pregnancy & 0 & 0 & 0 & 0 & 0 & 1 & 1(1) \\
weight & 9 & 1 & 0 & 0 & 0 & 0 & 10(7) & gender & 1 & 0 & 0 & 0 & 0 & 0 & 1(0) \\
location & 1 & 3 & 1 & 0 & 3 & 1 & 9(5) & reason & 1 & 0 & 0 & 0 & 0 & 0 & 1(1) \\
height & 3 & 1 & 0 & 0 & 1 & 0 & 5(4) & activity & 1 & 0 & 0 & 0 & 0 & 0 & 1(1) \\
duration & 1 & 1 & 2 & 0 & 1 & 0 & 5(5) & glucose & 0 & 1 & 0 & 0 & 0 & 0 & 1(1) \\
description & 1 & 1 & 0 & 0 & 2 & 1 & 5(5) & fat & 1 & 0 & 0 & 0 & 0 & 0 & 1(1) \\
note & 1 & 0 & 0 & 0 & 3 & 0 & 4(4) & account & 0 & 0 & 0 & 0 & 0 & 1 & 1(1) \\
date & 2 & 0 & 2 & 0 & 0 & 0 & 4(4) & doctor & 1 & 0 & 0 & 0 & 0 & 0 & 1(0) \\
age & 2 & 0 & 0 & 0 & 0 & 0 & 2(1) & registration & 0 & 0 & 1 & 0 & 0 & 0 & 1(1) \\
medication & 2 & 0 & 0 & 0 & 0 & 0 & 2(1) & \textbf{Sum} & \textbf{88} & \textbf{22} & \textbf{31} & \textbf{1} & \textbf{13} & \textbf{5} & \textbf{160(92)} \\
\bottomrule[1.5pt]
\end{tabular}
}
\label{Tab-Sec4-ACPInfo}
\vspace{-20pt}
\end{table*}

\subsubsection{Case Study of the Incomplete Privacy Policy}

As shown in Fig. \ref{Fig-Sec4-Incomplete},  the privacy policy of app \textit{com.bodyyouwant} is attached with three labels, i.e., \textit{Data Collection}, \textit{Data Usage}, and \textit{Contact Information}. It does not declare that it will ask for the users' consent before collecting user information, indicating that the collection process is non-transparent to the users. Furthermore, there is no user right described in the policy, indicating that the user does not know what they can do with the collected data. Such incomplete privacy policy is not clear to the app users and violates the GDPR. 
However,  \mytool{} is significant to guide app developers to provide complete policy and help app users quickly understand the semantics of important notices in the expatiatory policy.


\noindent\fbox{
	\parbox{0.95\linewidth}{
		\textbf{Answer to RQ1}:
		For the 796 mHealth apps, 189 (23.7\%) of them do not provide complete privacy policies.
		The incomplete privacy policy violates the GDPR and poses privacy issues for app users in the real world. It is imperative for app developers to complete existing incomplete privacy policies.
	}
}

\subsection{RQ 2: Do the mHealth apps declare all the collected PHI
	in their privacy policies? }
\label{subsec:evaBehav}

%

To answer this research question, we only focus on the 253 mHealth apps that require users to input PHI.

\subsubsection{Result of Inconsistent PHI Collection}
After the data-flow analysis, 59 of the 253 apps collect the user input PHI by storing them with six different channels.
To investigate the reasons for the other 194 apps that require the PHI input but have no PHI collection, we randomly select 20 apps from the 194 apps and  manually analyze the app code. We find that there are two main reasons.
First, most of the input PHI is only used to perform calculations such as health state, and the mHealth apps will not store the PHI with the six channels. Second, PHI collections might be missing due to the limitations of the existing data-flow analysis techniques.
For example, the Android annotation technique~\cite{annotation} makes \mytool{} fail to link the \textit{findViewById()} with the widget objects.
We will discuss this limitation in Section \ref{sec:discussion}.

We use the terms \textit{Log}, \textit{File}, \textit{Network}, \textit{SMS}, \textit{Database} and \textit{Content} to denote each data storage method, respectively. The frequency distribution of the collected PHI is listed in Table \ref{Tab-Sec4-ACPInfo}. There are 160 PHI collection behaviors for the 59 apps, in which name, email, and password are the most common ones.
Among the six data storage methods, \textit{Log} is the most common method since 55\% of the collected PHI is written into logs, \textit{SMS} is the least-used method because there is only one phone number collection behavior via sending messages.

After the construction of $DCP$ by tagging the declared PHI in the \textit{Data Collection} and \textit{Data Usage} sentences, we match them with the collected PHI in $ACP$.  The results show that 46 apps  have collected more PHI than what they declared in their privacy policies. The numbers of the non-declared PHI are listed in the brackets of the \textbf{Total} column in Table \ref{Tab-Sec4-ACPInfo}. For example, 21 apps collect the password of the users, but 16 of them do not declare the password collection in their privacy policies. In total, 92 inconsistent behaviors are discovered for the 46 apps. We observe that the app developers prefer to declare the common data, such as name, email, password, and phone number. However, for the uncommon but important data, such as duration, description, and date, they do not declare such data collection behaviors in the privacy policy. The inconsistent data collections violate the GDPR  and  mislead the app users.  Therefore, we want to alert the app developers to provide consistent privacy policy by conducting the inconsistent behavior analysis.



\subsubsection{False Positives and False Negatives}
 We further investigate the false positives and false negatives for inconsistent behavior analysis. For the false positive analysis, we initially   read all privacy policies  of the 46 mHealth apps that contain inconsistent behaviors. Then we check whether each inconsistent behavior is correct. The results show that among the 46 apps, no false positive is found. For the false negative analysis, since there is no ground truth for inconsistent behavior discovering, we manually analyze the 13 mHealth apps with no inconsistent behaviors to determine whether our approach results in false negatives. The results show that all the collected PHI has been declared in their privacy policies. Therefore, there is no false negative.

 \begin{figure}[t]
 	\centering
	\includegraphics[width=0.8\linewidth]{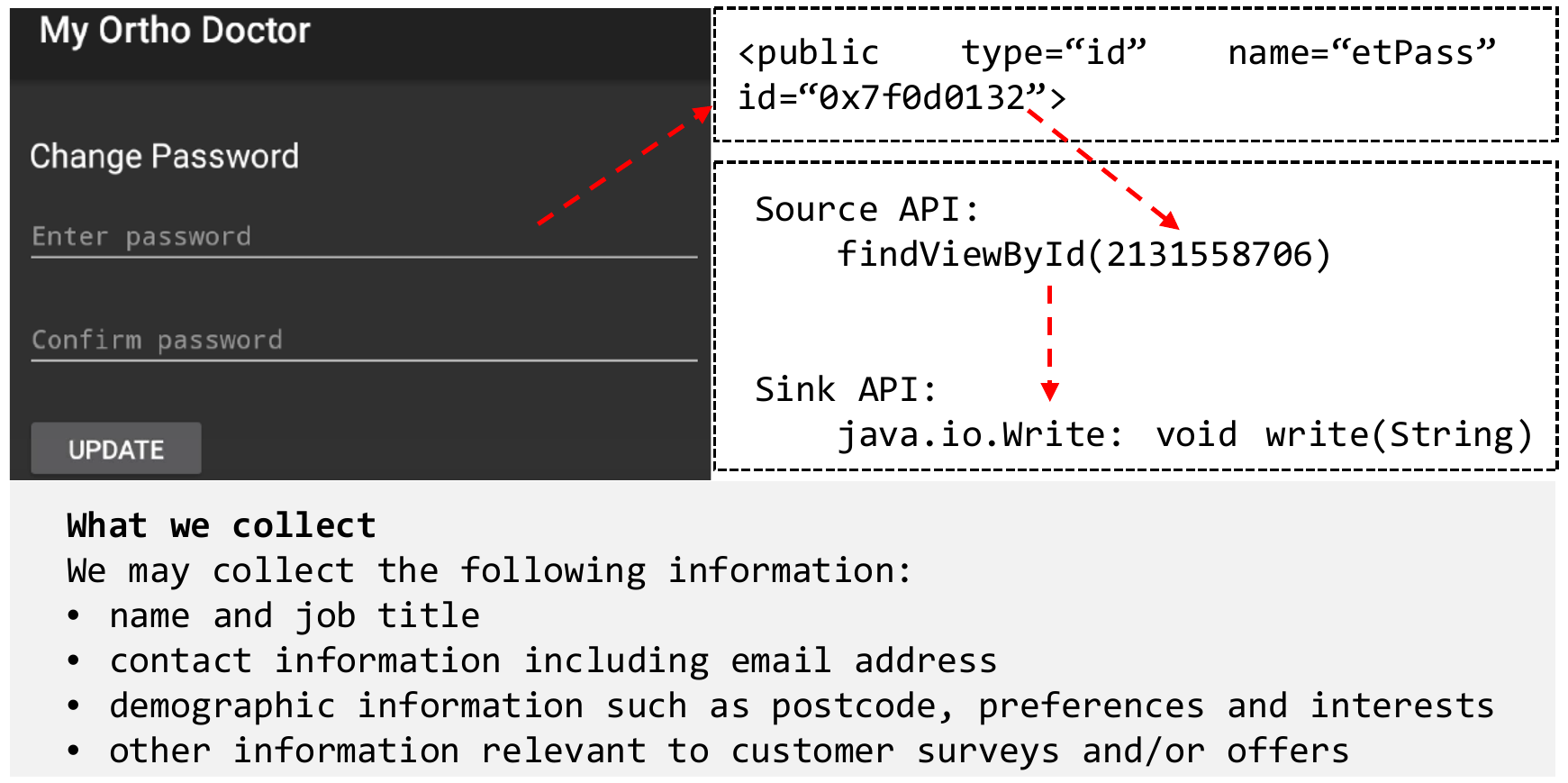}
	\caption{An inconsistent behavior of app \textit{com.app.app889ee1ec94b3}}
	\label{Fig-Sec4-Inconsistent}
	\vspace{-25pt}
\end{figure}

\vspace{-3pt}
 \subsubsection{Case Study of the Inconsistent Behavior}
 Furthermore, we conduct a case study of inconsistent behavior, which is illustrated in Fig. \ref{Fig-Sec4-Inconsistent}. The app provides the functionality of changing the password for app users. By analyzing the  attributes of \textit{EditText} widget, we obtain its hint value ``Enter password'' and \textit{id} ``0x7f0d0132''. Note that the hexadecimal number ``0x7f0d0132'' is equal to the decimal number ``2131558706''. After that, based on the data-flow analysis, the input password is written in a file through the sink API \textit{java.io.Write: write()}. The writing of input password to file poses great threats to app users. Even worse, the app developers do not declare that they collect password in their provided privacy policy, which violates the data minimization requirement of GDPR. Note that although they declare that other information relevant to the users is collected, this kind of vague description is also not transparent to app users and it should be specified clearly.


\noindent\fbox{
	\parbox{0.95\linewidth}{
		\textbf{Answer to RQ2}:
		Among the analyzed 253 apps, 59 apps collect PHI via different methods. However, 46 (77.9\%) of them  contain at least one inconsistent collection behavior.
		The app developers should not use vague descriptions about data collection, which might cause the inconsistent data collection behavior that poses great threats to the app users and seriously violates the GDPR.
	}
}


\subsection{RQ 3: Do the mHealth apps implement reasonable
	measures to ensure the transmission security of
	collected PHI?}
\label{subsec:evaSecure}

To identify whether the collected PHI for the 59 apps is protected with reasonable measures such as data encryption or SSL,
we analyze the correctness of encryption usage for the 49 apps that collect information via five methods (i.e., \textit{Log}, \textit{File}, \textit{SMS}, \textit{Database}, and \textit{Content}), and check the correctness of SSL usage for the other 10 apps that collect information through \textit{Network}.

\subsubsection{Result of Encryption and SSL Analysis}
After the detection of encryption function, we observe that only 2 apps (i.e., \textit{com.ysp.l30band} and \textit{kalcare.dsc}) have adopted system-provided encryption API calls to protect collected data. For exmaple,  \textit{com.ysp.l30band} encrypts the input password and email with \textit{MessageDigest.digest()} API call. By tracking the argument of the \textit{MessageDigest.getInstance()} API call we observe that both the password and the email are encrypted with MD5 algorithm, which breaks the security rule ``Do not use MD5 or SHA-1 algorithms for encryption.'' For the incorrectness encryption usage of app \textit{kalcare.dsc}, we leave it as a case study later.

Then, by applying \textsc{MalloDroid} \cite{fahl2012eve} on the 10 apps storing collected PHI through networks, we observe that
4 apps do not use SSL. For the other  6 apps that adopt SSL, each of them contains at least one SSL misuse such as trusting all certificates or allowing all hostnames. For example, the app called \textit{com.pumapumatrac} makes a blank implementation of  the \textit{TrustManager}  interface  so that it will trust all the server certificates (regardless of who signed it, what is the CN etc.). Furthermore, it even requires the use of \textit{SSLSocketFactory.ALLOW\_ALL\_HOSTNAME\_VERIFIER}. As a result, hostname verification  should take place when establishing an SSL connection is disabled.


\subsubsection{False Positives and False Negatives}
We also investigate the false positives and false negatives for insecure data transmission identification. Since our results demonstrate that the data transmission of all the 59 mHealth apps is insecure, there is no false negative. To evaluate whether there exist any false positives, we manually check all the data flow information of the collected PHI for the apps that do not use encryption function or SSL. We observe that all the PHI is stored in plaintext. Furthermore, we manually check the 6 apps that use SSL and do not find any false positives.

\subsubsection{Case Study of the Insecure Data Transmission}
\begin{figure}[t]
	\centering
	\includegraphics[width=0.8\linewidth]{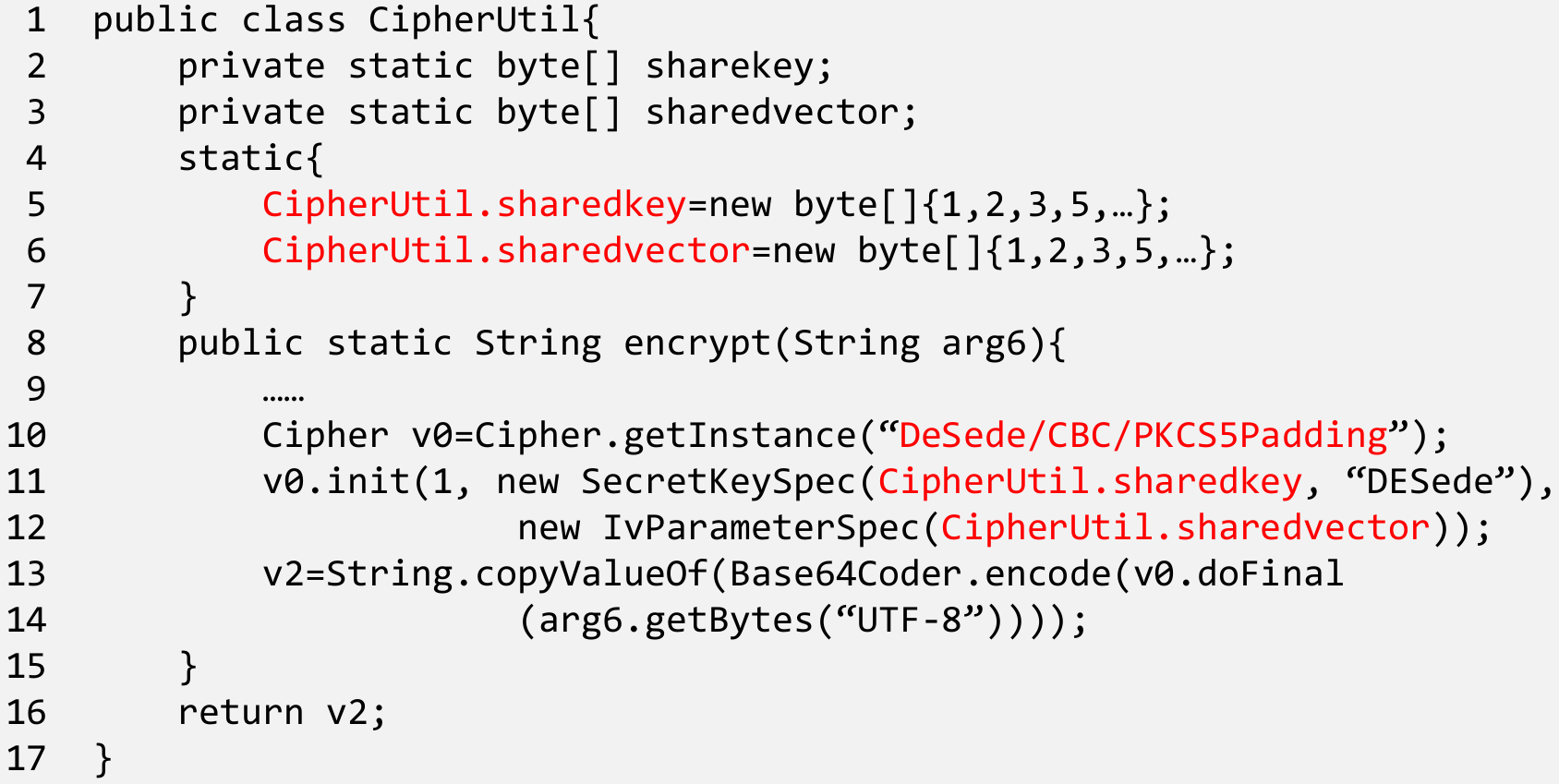}
	\caption{Code snippets of encrypting data with the \textit{doFinal()} API call in app \textit{kalcare.dsc}. }
	\label{Fig-Sec4-Encryption}
	\vspace{-20pt}
\end{figure}

Finally, we take the encryption detail of app \textit{kalcare.dsc} as a case study. \textit{kalcare.dsc} encrypts four kinds of PHI, including name, phone, email, and password. The collected name and phone are encrypted with \textit{MessageDigest.digest()} API call and the collected email and password are encrypted with \textit{Cipher.doFinal()} API call. By tracking the argument of \textit{MessageDigest.digest()} we observe that SHA-256 algorithm is used. For the use of \textit{Cipher.doFinal()} API call, we further check its compliance with the security rules.  The code snippets of encrypting email and password with the \textit{doFinal()} API call are presented in Fig. \ref{Fig-Sec4-Encryption}. The argument of \textit{Cipher.getInstance()} is ``DESede/CBC/PKCS5Padding,'' which indicates that the adopted encryption algorithm is DESede and the encryption scheme is CBC mode. Thus, this app obeys the rule ``Do not use ECB mode for encryption'' but does not follow the rule ``Do not use a constant IV or constant keys for encryption'' as both the arguments of \textit{SecretKeySpec()} and \textit{IvParameterSpec()} are static constants listed in line 5 and line 6, which  would cause data more subject to attacks.

\noindent\fbox{
	\parbox{0.9\linewidth}{
		\textbf{Answer to RQ3}:
		For the 59 mHealth apps that collect user input PHI, only 8 of them  try to ensure the PHI transmission security with reasonable measures, i.e., 2 apps protect data with encryption algorithms and 6 apps adopt SSL protocol. However, all of the 8 apps contain at least one kind of encryption or SSL misuse. The security of the collected PHI data has not been heeded enough, which would cause serious data breaches.
	}
}

\section{Discussions and Threats to Validity}
\label{sec:discussion}


\subsection{Lessons Learned from Results}

Providing transparent and accessible privacy policy about the personal data is the essential requirement for organizations  that fall under the scope of GDPR. 
 However, according to our evaluation result, 189 (23.7\%) do not provide complete privacy policies. The main reason is that GDPR introduces several new privacy requirements compared with old regulations. For example,  one new requirement is that apps must acquire the user's active and informed consent before any personal data is collected (i.e., \textit{User Consent} notice in this work). However, up to now, many apps would assume that a user's decision to proceed with app registration and use is equivalent to having the user's consent to collect data. Lacking any notice introduced in our work would not meet the transparency requirement of GDPR. By using our tool, app developers can discover the missing notices and complement them. For app users, they can quickly understand the semantics for the most important data processing related content, so well as their rights when using the app.


Data minimization is another essential requirement of GDPR. Data processing should only use as much data as is required to successfully accomplish a given task. However, in our work, we find that 46 apps contain at least one inconsistent data collection behavior. We manually check the policies and find there are two reasons: First, 59 inconsistent behaviors occurred in 26 apps are caused by the vague description such as ``other information''; Second, the other  33 inconsistent behaviors occurred in the other 20 apps are mainly caused by the app developers'  intentional ignorance (e.g., they think such data is not important, or they want to collect such data without informing app users). To mitigate the occurrence of inconsistent data collection behaviors, by leveraging our tool, the app developers can first list all their collected data. Then they can specify such collected data in their privacy policy clearly. For app users, they can have a clear understanding of which personal data are collected by app developers and how they are processed in the app. 



Confidentiality is the only principle that deals explicitly with security in GDPR. Meanwhile, it is  the most concerned one by users since there are more and more data  breach events in recent years. Based on our evaluation results, although most apps declare that they would implement reasonable measures to keep the data secure, only 8 apps try to adopt security measures, and all of them contain at least one kind of  misuse. The evaluation results indicate that the app users' data might be leaked in high probability, which is amazing to us. By leveraging our tool, we strongly suggest the  app developers enhance their data protection methods by checking with the standard rules.


\subsection{Threats to Validity}
\noindent\textbf{Data-flow Analysis.} Even we use the popular static analysis tools, including  \textsc{FlowDroid}, \textsc{IccTA}, and \textsc{EdgeMiner}, to tract the data flow, 
there still exist false negatives, which might cause the missing of PHI collection.
The existing of false negatives further unveils that the status quo of GDPR compliance violations in Android apps is worse than what we demonstrate.
Combining with dynamic analysis techniques \cite{xue2017malton, tam2015copperdroid} is a promising way to  improve the accuracy of PHI collection further, and we leave it as our future work.

\noindent\textbf{Self-implemented Encryption Detection.} For the encryption analysis, our approach would fail if the self-implemented encryption function is applied since we only rely on the study of system-provided encryption algorithms. A promising way to address this limitation is to compare the data entropy before and after the invocation of possible encryption function while the app is running  \cite{wood2017cleartext}. If the data entropy is much higher after the function invocation, then the data might be encrypted in the corresponding function.


\noindent\textbf{Ground Truth Dataset:} We use a triple module redundancy approach when preparing ground truth dataset. However, if three authors fail to achieve a consistency, we would not add the sample into our dataset. The lacking of such sentences in training set would affect the classifier performance when dealing the similar sentences.

\section{Related work}
\label{sec:Relatedwork}
%
\noindent{\bf Privacy Policy Analysis.}
Several studies focus on the privacy policy analysis in recent years~\cite{harkous2018polisis,andow2019policylint,kafali2017good,zimmeck2017automated}. 
Slavin \textit{et al.} \cite{slavin2016toward} proposed a semi-automated framework for detecting the violations based on a privacy-policy-phrase ontology and a collection of mappings from API calls to policy phrases.  Yu \textit{et al.} \cite{yu2016can} proposed \textsc{PPChecker} that focuses on system-managed data and  identify three kinds of problems in the privacy policy. 
The most related work is proposed by Wang \textit{et al.} \cite{wang2018guileak}, who automatically detect
privacy leaks of user-entered data for a given Android app and determines whether such leakage may violate the app's privacy policy claims. 
There are three main differences between our work and the above studies: 1) We combine the analysis of mHealth apps with the GDPR while \cite{slavin2016toward, zimmeck2017automated, yu2016can, wang2018guileak} do not consider. 2) We focus on the PHI input by the users on GUI rather than the system-managed data analyzed by \cite{slavin2016toward, zimmeck2017automated, yu2016can}. 3) We further investigate the transmission security of collected data while \cite{wang2018guileak} does not.


\noindent{\bf GUI Analysis.}
A few studies are focusing on the analysis of GUI \cite{huang2014asdroid, mulliner2014hidden, andow2017uiref, huang2015supor, nan2015uipicker, yang2015static, azim2013targeted}. The most related works are \textsc{UIPicker} \cite{nan2015uipicker}, \textsc{SUPOR} \cite{huang2015supor}, \textsc{UiRef} \cite{andow2017uiref} and \textsc{GUILeak} \cite{wang2018guileak}, the goals of which are identifying the sensitive user input information on the GUI. \textsc{UIPicker}  \cite{nan2015uipicker} and \textsc{GUILeak} \cite{wang2018guileak} use sibling relationships in layouts to find the associated labels and input widgets. However, in practice, sibling relationships do not accurately gauge proximity. Both \textsc{SUPOR} \cite{huang2015supor} and \textsc{UiRef} \cite{andow2017uiref} select the optimal label by calculating the distances between the labels and the input widgets based on the positions displayed on the screen. 


\noindent{\bf GDPR Compliance Checking.} 
Several recent works focus on the GDPR compliance checking~\cite{torre2019using,palmirani2018modelling,ayala2018grace,tom2018conceptual}. However, their methodologies are quite different from ours. Torre \textit{et. al}~\cite{torre2019using} proposed a model-based GDPR compliance analysis solution using unified modeling language (UML) and object constraint language (OCL). 
Palmirani and Governatori~\cite{palmirani2018modelling} presented a proof-of-concept applied to the GDPR domain,
with the aim to detect infringements of privacy compulsory norms or to prevent
possible violations using BPMN and Regorous engine. 
 These existing
 approaches conduct compliance checking from the perspective of modeling. We go beyond the above approaches by 
  transforming the GDPR requirements into specific regulations and combine with program analysis techniques so that we can conduct a fine-grained empirical evaluation on real mHealth apps.

\section{Conclusion}\label{sec:conclution}

We develop a system called \mytool{} based on existing techniques and conduct the first systematic investigation on automatically detecting the compliance violations between the GDPR and mHealth apps. 
The experimental results on  796 real mHealth apps reveal that most of the apps are not compliant with the GDPR, which would raise the awareness of the privacy protection for the mHealth app users and developers. We hope that our tool can  help developers identify and fix problems before releasing apps and allow users to make informed decisions about the apps that they use.



\section*{Acknowledgment}
This work was supported by National Key R\&D Program of China
(2016YFB1000903), National Natural Science Foundation of China
(61902306, 61632015, 61772408, U1766215, 61721002, 61532015, 61833015),
Ministry of Education Innovation Research Team (IRT\_17R86), and
China Postdoctoral Science Foundation.

\bibliographystyle{IEEEtran}
\bibliography{IEEEabrv,appgdpr}

\end{document}